 \documentclass[final,5p,times,twocolumn,authoryear]{elsarticle}

\usepackage{amssymb}
\usepackage{lipsum}

\journal{High Energy Astrophysics}

\usepackage{hyperref}
\usepackage{amsmath}
\usepackage{derivative}
\usepackage[caption=false]{subfig}
\usepackage{soul, color}

\begin{document}

\setstcolor{red}

\begin{frontmatter}



\title{Modelling the energy dependent X-ray variability of Mrk 335}


\author[first]{K. Akhila} \ead{akhilasadanandan@gmail.com}
\author[second]{Ranjeev Misra}
\author[third]{Rathin Sarma}
\author[fourth,fifth]{Savithri H. Ezhikode}
\author[first]{K. Jeena} \ead{jeenakarunakaran@gmail.com}

\affiliation[first]{organization={Department of Physics, Providence Women's College (Autonomous)},
            addressline={University of Calicut},
            postcode={673009}, 
            state={Kerala},
            country={India}}
\affiliation[second]{organization={Inter-University Centre for Astronomy and Astrophysics (IUCAA)}, 
                    addressline={PB No.4, Ganeshkhind},
                    city={Pune},
                    postcode={411007},
                    country={India}}
\affiliation[third]{organization={Department of Physics, Rabindranath Tagore University},
                    city={Hojai},
                    postcode={782435},
                    state={Assam},
                    country={India}}
\affiliation[fourth]{organization={St. Francis de Sales College (Autonomous)},
                    addressline={Electronics City},
                    city={Bengaluru},
                    postcode={560100},
                    country={India}}
\affiliation[fifth]{organization={Department of Physics and Electronics, CHRIST (Deemed to be University)},
                    city={Bangalore},
                    postcode={560029},
                    country={India}}

\begin{abstract}
We present a technique which predicts the energy dependent fractional r.m.s for linear correlated variations of a pair of spectral parameters and apply it to an \textit{XMM-Newton} observation of Mrk 335. The broadband X-ray spectrum can be interpreted as a patchy absorber partially covering the primary emission, a warm and hot coronal emission or a relativistically blurred reflection along with the primary emission. The fractional r.m.s has a non-monotonic behavior with energy for segments of lengths  3 and 6 ksecs. For each spectral model, we consider every pair of spectral parameters and fit the predicted r.m.s with the observed ones, to get the pair which provides the best fit. We find that a variation in at least two parameters is required for all spectral interpretations. For both time segments, variations in the covering fraction of the absorber and the primary power law index gives the best result for the partial covering model, while a variation in the normalization and spectral index of the warm component gives the best fit in the two corona interpretation. For the reflection model, the best fit parameters are different for the two time segment lengths, and the results suggests that more than two parameters are required to explain the data. This, combined with the extreme values of emissivity index and reflection fraction parameters obtained from the spectral analysis, indicates that the blurred reflection model might not be a suitable explanation for the Mrk 335 spectrum. We discuss the results as well as the potential of the technique to be applied to other data sets of different AGN.
\end{abstract}




\end{frontmatter}




\section{Introduction} \label{sec:intro}

Studying the X-ray emission is one way to understand the physics of the inner mechanisms in active galactic nuclei (AGNs). A long known property of AGN emission is the X-ray flux variability. Long term variabilities, spanning from a few days to years, were first identified with the \textit{Ariel 5} and \textit{HEAO 1} missions (e.g., \citealt{marshall1981variability}). Short rapid variabilities of the order of thousands of seconds were identified much later through missions such as \textit{EXOSAT} (e.g., \citealt{mchardy1985x}). In the $0.1$ - $10$ keV band, flux variations are observed on timescales scanning a wide range from a few thousand seconds to years, having amplitude variations upto an order of magnitude \citep{mchardy1988exosat, green1993nature, turner1994x, turner1999x}. 

A multitude of studies have been conducted on the AGN X-ray flux variabilities \citep{lawrence1987low, lawrence1993x, nandra2001x, martin2003variability, markowitz2003long, soldi2014long, pryal2015search, komossa2016extremes, papadakis2024x, serafinelli2024investigating}. One way of quantifying the variability is through statistical tools such as excess variance and fractional rms variability amplitude ($F_{var}$ : \citealt{vaughan2003characterizing}). These quantities are estimated after removing the contribution from measurement errors and are hence an indicator of the intrinsic variability of the source.

The X-ray spectra of a typical Seyfert galaxy exhibits a power law emission along with the presence of broad and narrow emission lines \citep{fabian2000broad,tatum2012modeling}. At low energies ($<2$ keV) the flux from the power law emission is enhanced by another smooth component, known as the soft excess. It was first observed in the year 1985 by \citet{singh1985observations} in the Seyfert-1 galaxy Mrk 509 and by \citet{arnaud1985exosat} in Mkn 841 using \textit{HEAO} and \textit{EXOSAT} observations, respectively. The soft excess has since been observed in several Narrow Line Seyfert-1 (NLS1) galaxies \citep{boller1995soft, buhler1995rosat, middleton2007absorption, middei2019high, boller2023unraveling}. Soft excess refers to an additional emission in the soft X-ray energies of these AGNs, which cannot be explained by the low-energy extrapolation of their hard energy continuum \citep{done2007can, done2012intrinsic, middei2020soft}. Observations have revealed the presence of soft excess in both broad and narrow line Seyfert-1 galaxies, but with a stronger strength in NLS1 galaxies \citep{gliozzi2020soft}.

Several different theoretical models including the obscured model, warm coronal picture and a gravitational blurred reflection model have been used to explain the soft excess. Despite the crucial differences in their underlying physics, all these proposed models were quite successful in explaining the observed spectrum \citep{Sobolewska_2005, ding2022variability}. The partial covering model describes the excess as a `leakage' through a non-uniform absorber. The power law continuum is obscured by the absorber. The presence of such patchy clouds, introduce variations in spectral behavior, waning a fraction of the incident radiation. Both this obscured emission and direct primary emission contribute to the observed spectrum. The changes in spectral features may then be attributed to changing line-of-sight absorber, without requiring alterations in the primary emitter. \citep{weaver1993complex, tanaka2004partial, grupe2008xmm, reeves2008iron, gallo2011multi, gallo2015suzaku, iso2016origin, tripathi2019nature, parker2019nuclear, komossa2020lifting}. Several studies, including earlier ones like \citet{weaver1993complex} and \citet{turner1996complex}, use the partial covering model to explain the X-ray spectra of Seyfert galaxies \citep{chevallier2006role, middleton2007absorption}. In the warm thermal Comptonization picture, there is an optically thick warm corona which is distinct from the hot corona. The seed photons from the far ultraviolet end of the accretion disk undergo Compton upscattering by warm ($\sim0.1$ keV) electrons in the corona giving rise to heightened emissions at low X-ray energies. The hot corona is responsible for the power law emission at higher energies \citep{dewangan2007investigation, middleton2009re, done2012intrinsic, ezhikode2017determining, petrucci2018testing, kubota2018physical, ursini2020nustar}. A multi-epoch, multi-instrument analysis on the Seyfert-1 galaxy Zw 229.015 by \citet{tripathi2019nature} suggested that the smooth and featureless spectrum favors the Comptonization picture. \citet{garcia2019implications}, on the other hand, argue that the parameters required for thermal Comptonization are physically incompatible with the conditions of a standard corona. They put forward the relativistic reflection model as a better description of the observation. This gravitationally smeared ionised reflection component is another popular candidate for the origin of the soft excess \citep{miniutti2004light, crummy2006explanation, liebmann2018x}. It refers to the relativistic reflection from the innermost regions of the accretion disk. The X-ray continuum gets reflected from the accretion disk and is gravitationally blurred due to the close proximity of the black hole. Presence of soft reverberation lags also support the reflection model \citep{fabian2009broad, alston2020dynamic}. The reflection model has an additional advantage that it paves a way to understand the accretion mechanism in NLS1 galaxies \citep{reynolds2021observational}. However, a study by \citet{adegoke2017spectral} showed that both the models are statistically similar in explaining the soft X-ray excess. Another recent \textit{XMM-Newton} study on a sample of six NLS1 galaxies by \citet{yu2023xmm} also gives a similar result. They confirm the presence of a warm and optically thick corona in the Comptonization model while reflection picture is a better explanation for the behaviour of single source.

In this work, we present a technique to model the source's energy dependent X-ray variability and recognise the parameters causing the variability for all spectral models. We determine the observed fractional rms ($F_{var}$) as a function of energy from the source lightcurves extracted in multiple energy bands. The different spectral models may then be used to predict the fractional rms. The X-ray variability is attributed to a pair of spectral parameters, varying in correlation, from which, $F_{var}$ is then predicted and the parameter pair that best recreates the observed temporal variation is identified. We demonstrate the method by an investigation of the X-ray spectrum of Markarian 335 (Mrk 335).

Mrk 335 is one of the widely studied Seyfert-1 galaxies \citep{longinotti2007evidence, grupe2008xmm, tombesi2010evidence, liu2010difference, patrick2011iron, longinotti2013rise, de2013discovery, kara2013discovery, wilkins2015flaring, sarma2015relationship, chainakun2015simultaneous, epitropakis2016theoretical, chainakun2016relativistic, emmanoulopoulos2016search, choudhury2019testing, ingram2019public, ezhikode2021astrosat, liu2021systematic, liu2021observational, yang2022exploring, hancock2022x}. It harbours a supermassive black hole of mass $\sim2.6 \times 10^7M_\odot$ \citep{grier2011reverberation}. Its X-ray spectrum resembles that of a typical NLS1 galaxy. It exhibits a very strong soft excess and prominent iron lines \citep{o2007relativistic, larsson2008suzaku}. The soft excess in Mrk 335 was first observed in 1987 and has since been extensively observed \citep{pounds1987discovery, turner1988variability}. Mrk 335 has been observed to be variable in X-ray \citep{arevalo2008x, gallo2018eleven} and we make use of this variability to understand the origin of the soft excess in the source.

This paper is organised as follows. The details of observation and data reduction techniques employed are given in Section \ref{sec:data}. Section \ref{sec:spectrum} explains the different ways in which the source's X-ray spectrum was modelled. In Section \ref{sec:Fvar}, $F_{var}$ determination and rms spectrum are explained. In Section \ref{sec:Fpred}, the method used for predicting $F_{var}$ from theoretical models is detailed. Finally we discuss the results in Section \ref{sec:result}.

\section{Observations and Data Reduction} \label{sec:data}

ESA's \textit{XMM-Newton} Observatory \citep{jansen2001xmm} has observed Mrk 335 several times since 2000. The source was observed to have fallen into a state of very low flux in 2007 \citep{grupe2007discovery, longinotti2008seyfert} and has continued to remain so with occasional flaring episodes. For our analysis, we selected a $\sim133$ ks long observation made in 2006 (Obs Id 0306870101, PI : Nandra) while the source was still in its brightest stage and showed large variability.

We consider the epic-pn camera \citep{struder2001european} data alone for the analysis. The data was processed using the Science Analysis Software \textsc{sas}-20.0.0. Following the standard procedures, the data was first filtered to remove the high background flares. \textsc{sas} tool \textsc{tabgtigen} was used to set the threshold value for the count rates and a file containing the good time intervals (GTI) was created. Data with the conditions PATTERN$\leq$4 and FLAG==0 only were selected. The task \textsc{epatplot} was used to verify that the observation was free from pile-up. X-ray spectrum was extracted from a central circular region of radius 35". The \textsc{sas} tasks \textsc{rmfgen} and \textsc{arfgen} were used to generate the response matrix and auxiliary response files respectively. The task \textsc{specgroup} was used and the spectrum was grouped to have a maximum of three bins per resolution FWHM, with a minimum of 30 counts in every bin. Spectral fitting was done using the \textsc{xspec}-12.12.0 tool \citep{arnaud1996xspec} of \textsc{heasoft}. Lightcurves were extracted with a time bin size of $100$ s from a central source region of size 35". \textsc{epiclccorr} was used to subtract background lightcurves extracted from a region of the same size and to account for dead time and exposure variations. 

\section{Spectral Analysis} \label{sec:spectrum}

\begin{figure}
    \centering
    \includegraphics[width=\linewidth]{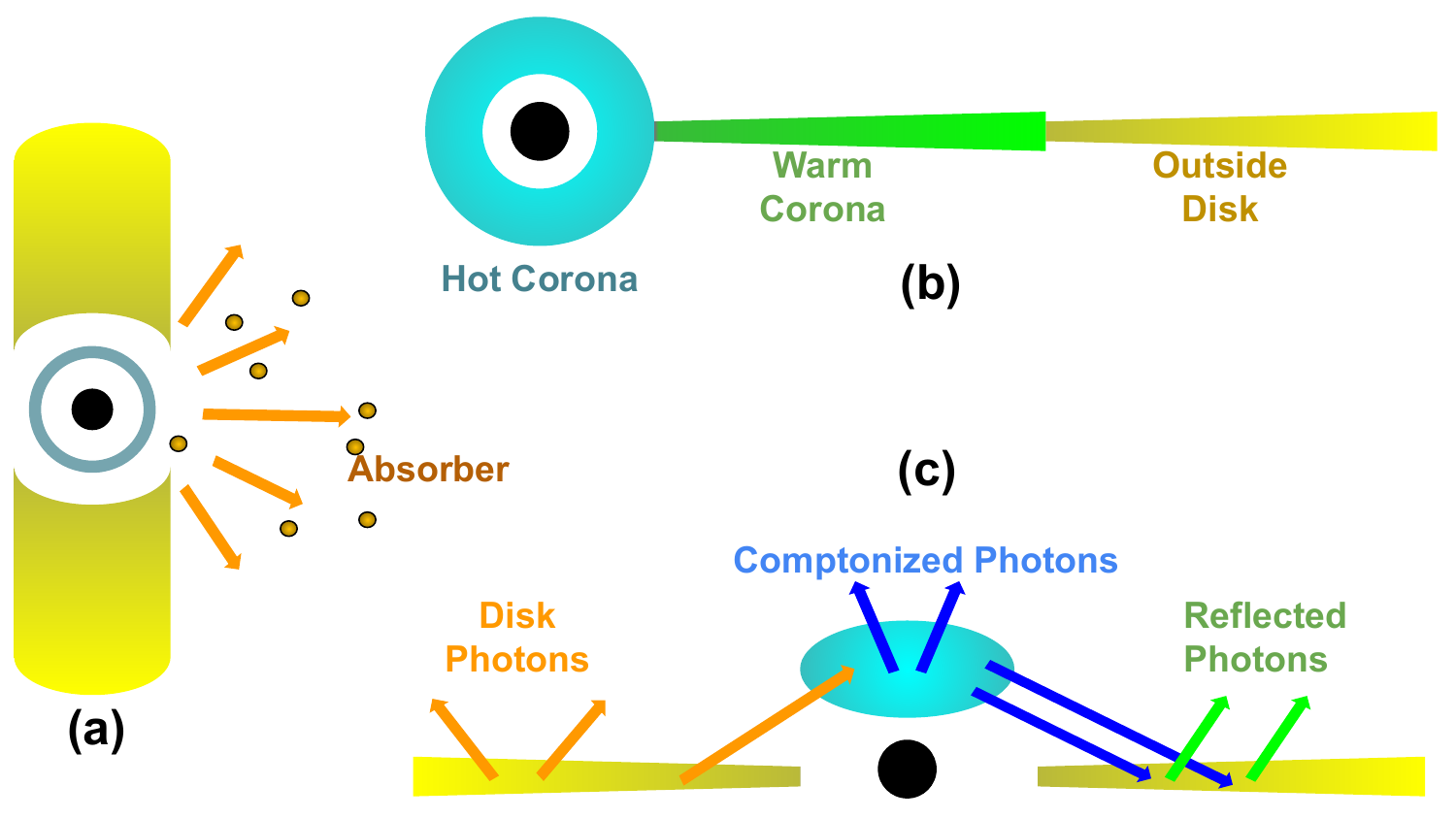}
    \caption{(a) : Patchy absorber partially covering the primary emission. (b) : Thermal Comptonization model geometry, where the disk (yellow) photons enter the warm Comptonization region (green) and then a fraction of these photons from warm corona enter the hot corona (blue). (c) : Blurred reflection model, wherein the corona (blue) illuminates a portion of the accretion disk (yellow). The coronal geometry maybe of different flavors like the lamppost model, spheroidal or toroidal corona or the sandwich model.}
    \label{fig:illust}
\end{figure}

Spectral fitting was done for the epic-pn X-ray spectrum in the $0.3 - 10.0$ keV energy band. The \textsc{xspec} tool uses the $\chi^2$ statistic to perform the fit. We obtain the errors from \textsc{xspec}, which gives their $90\%$ confidence interval. The X-ray spectrum was a power law continuum overlaid by emission lines. Prominent emission lines were seen in the iron band at $6.4 - 7.0$ keV and were fit with Gaussian (\textit{zgauss}) components. These correspond to the neutral Fe $K\alpha$ emission line at $\sim6.4$ keV and ionised Fe xxvi line at $\sim7.0$ keV \citep{o2007relativistic}. Cosmological redshift for the source was taken to be $z = 0.02578$ \citep{Huchra_1999} and the Galactic absorption ($N_H$) was fixed at $3.56\times10^{20}cm^{-2}$ \citep{kalberla2005leiden}. Previous studies have revealed a significant presence of warm absorber zones in the source spectra \citep{gallo2013blurred}. Hence we added two absorption edges at low energies which improved the spectral fitting. The observed soft X-ray excess of the source has been a subject of detailed analysis \citep{grupe2008xmm, brightman2011xmm, zhong2013magnetic, parker2014nustar, gallo2015suzaku, wilkins2015driving, keek2016revealing, boissay2016hard}. These analyses demonstrate that the broadband X-ray spectrum can be described by a partial covering model, as a Compton upscattering of the seed photons or by a relativistic reflection model. We fit the spectrum accordingly.

\subsection{Model-I : Partial Covering Model}
In the partial covering scenario, the intrinsic continuum emission of the source is modified by a partially covering absorber. In this model, the spectrum is described with the absorber reshaping the underlying single power law. We follow the spectral model prescribed in \citet{grupe2008xmm}, for the same observation. Partial covering of the primary emission is achieved by applying the \textit{zpcfabs} component to power law. This absorber model gives a better fit on adding Gaussian components of widths $0.25$ keV, that mimic line emission for the soft emission lines. Additionally, we added an absorption edge at $0.42$ keV, which improved the fit. The two soft energy Gaussians are added at energies $\sim0.1$ keV and $\sim0.8$ keV. We obtained a continuum power law index of $2.31$. The column density for the partial covering absorber is $N_{H,pc}$ = $45.77\times10^{22}$ $cm^{-1}$ with a covering fraction of $0.46$. We obtained a fit statistic of $\chi^2$/d.o.f. = $302.3/163$. While this value is on the higher side, we note that, following the data grouping adopted in \citet{grupe2008xmm}, we obtained a $\chi^2$/d.o.f = $917.4/700$, closer to the reduced $\chi^2$ of $1.2$ in \citet{grupe2008xmm}. The spectral fitting plot is shown in Figure \ref{fig:fig_pcf} and the parameters of fit are listed in Table \ref{tab:tab_pcf}.

\subsection{Model-II : Thermal Comptonization}
Here, the X-ray continuum was modeled as Comptonized emission of the seed photon spectrum. The seed photons are Compton upscattered by the hot electrons in the corona resulting in an enhanced emission at low energies. As already discussed in Section \ref{sec:intro}, in this model, the soft X-ray excess is considered to be due to the warm coronal component \citep{porquet2018deep, middei2018multi, noda2018explaining, matzeu2020first}. Such a two component model can have a geometry where the two occupy physically different regions. Here, the warm corona is connected to the accretion disk, while the hot corona lies beyond (Figure \ref{fig:illust}). We consider the case where the UV disk photons enter the warm Comptonizing region first and then the photons from the warm coronal medium enter the hot corona \citep{ezhikode2016uv, kubota2018physical, zoghbi2023measuring}. Warm corona Comptonizes the disk photons, giving the soft emission and a fraction of these photons are further scattered by the hot corona and emit in the hard X-ray energies. We use two \textsc{xspec} models, namely \textit{simpl} and \textit{nthComp}, to represent the hot and warm components, respectively. The disk photons enter the warm corona represented by \textit{nthComp} emitting in the soft region, while a fraction of this is further upscattered by the hot corona represented by \textit{simpl}, an empirical model for the Comptonization of a fraction of seed photons to a higher temperature \citep{steiner2009simple}. The seed photons were considered to be blackbody in shape (inp\_type=0). The spectral fitting plot is shown in Figure \ref{fig:fig_nthc}. We obtained a photon index ($\Gamma_{simpl}$) value of $\sim1.82$. The seed photon temperature obtained was $\sim5.98\times10^{-2}$ keV. The spectral fit gives a value of $0.98$ keV for the electron temperature. While we were unable to obtain a constraint on its upper limit, the parameter had a lower limit of $0.69$ keV. A complete list of the parameters of fit is given in Table \ref{tab:tab_nthc}.

\subsection{Model-III : Relativistic Reflection}
\begin{figure}
\centering 
\subfloat{%
  \includegraphics[width=0.9\linewidth]{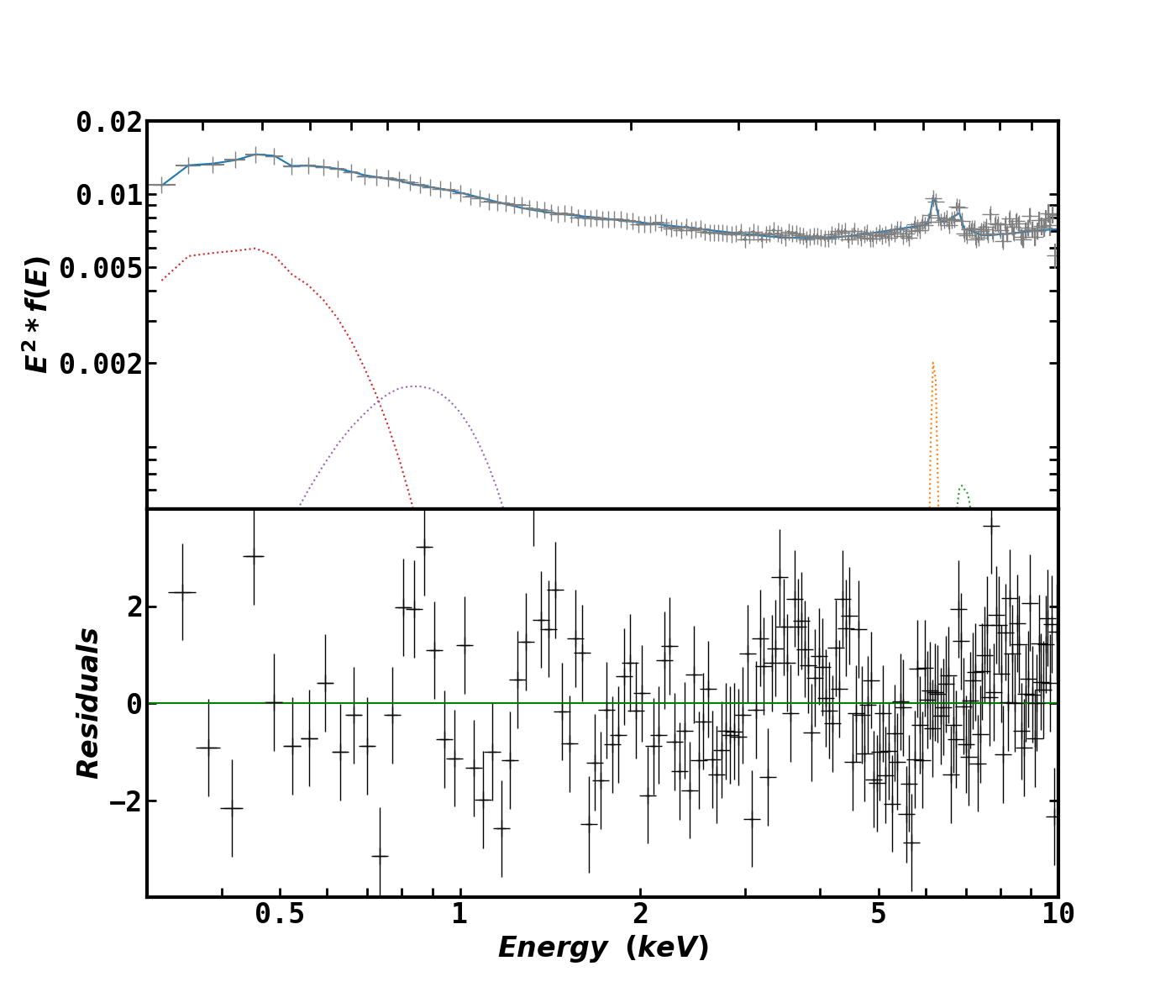}
  \label{fig:fig_pcf}%
}\qquad
\subfloat{%
  \includegraphics[width=0.9\linewidth]{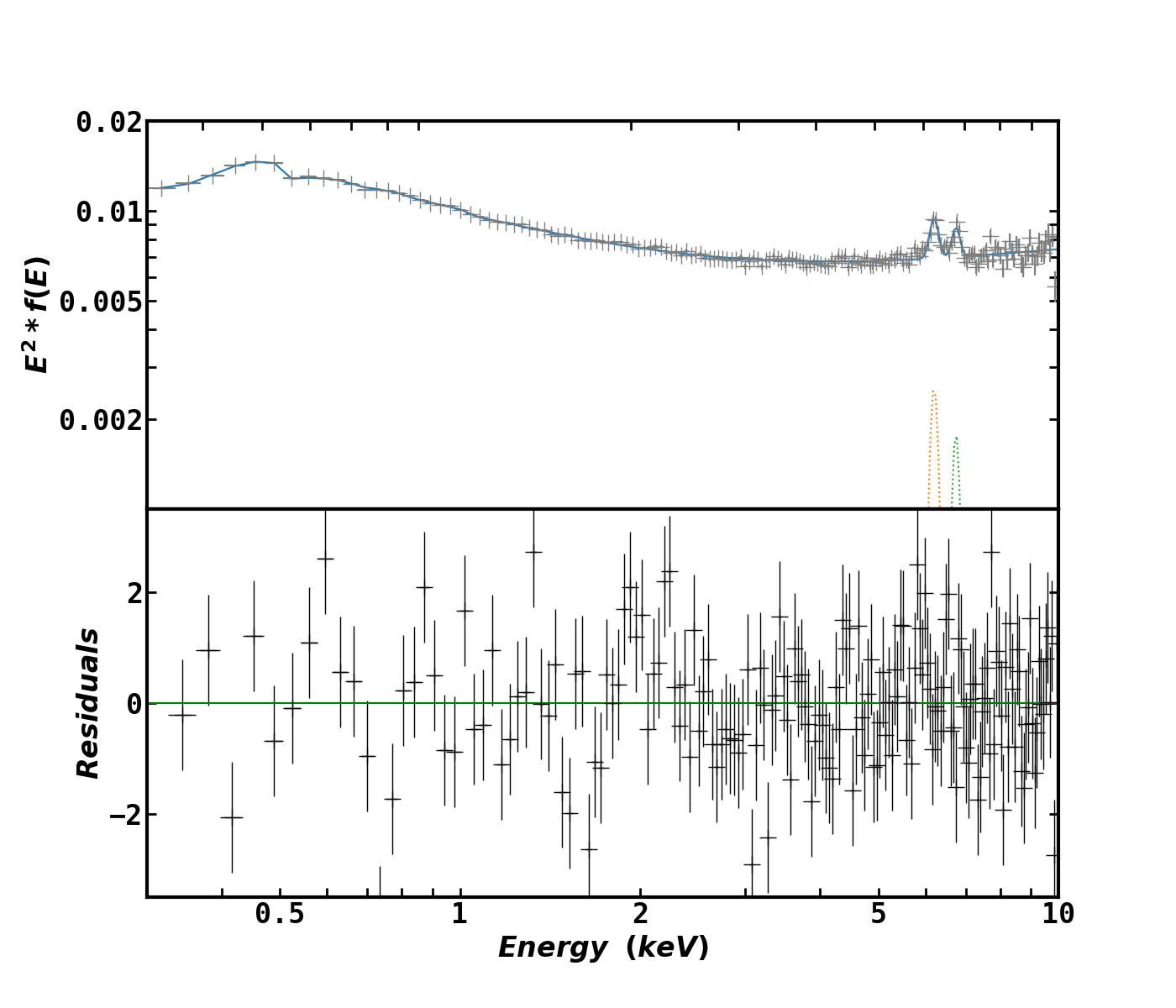}%
  \label{fig:fig_nthc}%
}\qquad
\subfloat{%
  \includegraphics[width=0.9\linewidth]{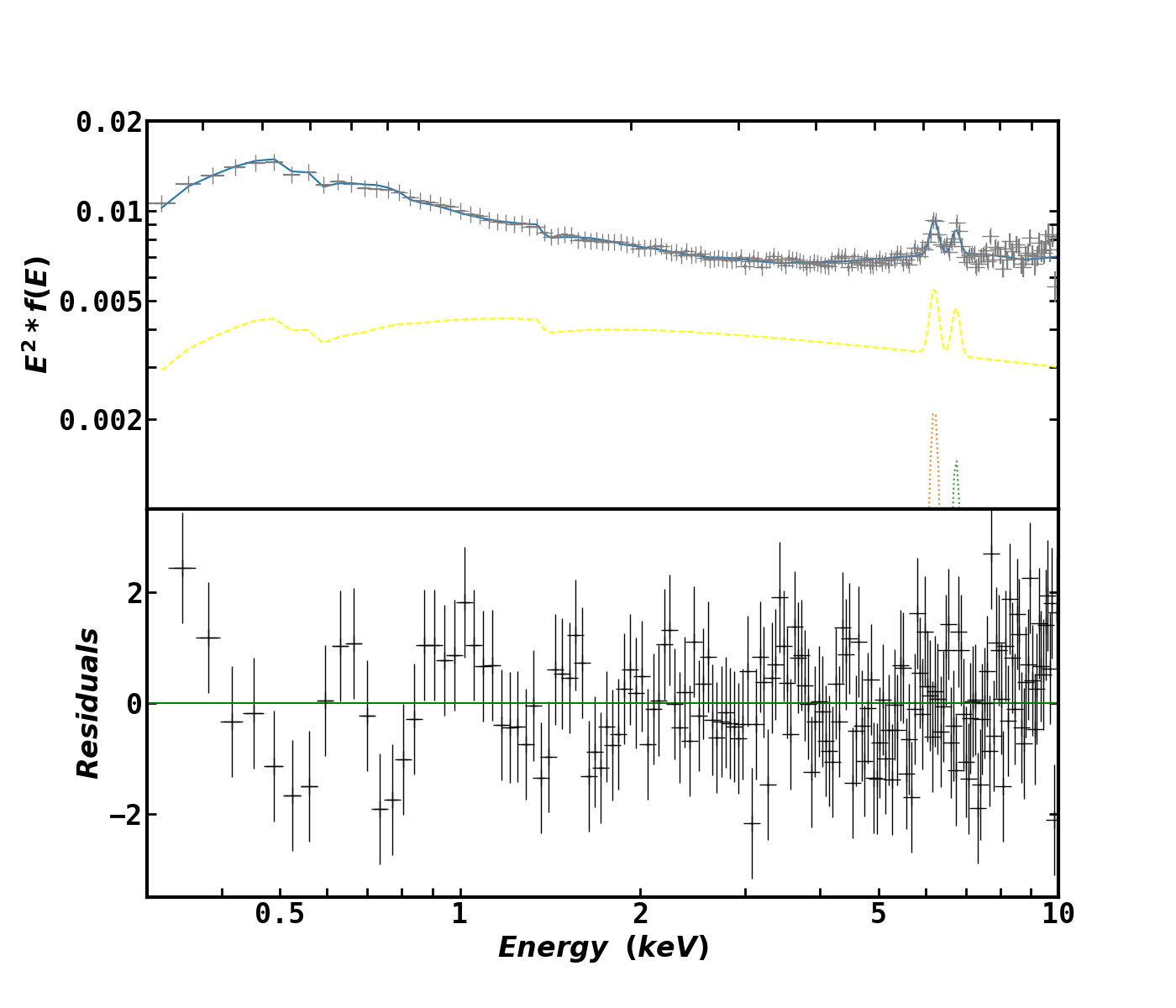}%
  \label{fig:fig_relx}%
}
\caption{X-ray spectral fitting of epic-pn observation in the energy range $0.3-10.0$ keV. Spectral fitting using the three models namely, partial covering model (Model-I, top row), the thermal Comptonization model (Model-II, middle row) and the blurred reflection model (Model-III, bottom row), respectively are shown. In all figures, dotted lines represent the individual additive components in the model.}
\end{figure}

In this case, the spectrum was modelled as a standard relativistic reflection model. Blurred reflection features arise in the spectrum when the `cold' accretion disk is illuminated by the `hot' corona. Since the disk efficiently radiates, it is colder than the coronal material. The geometry of this hot corona is still not properly understood, however, some possibilities discussed in literature are the lamppost model, spheroidal or toroidal corona and the sandwich model. A fraction of the photons from the corona can illuminate the disk, giving rise to the reflection spectrum \citep{bambi2024towards}. The \textit{relxill} code \citep{dauser2014role, garcia2014improved, dauser2016normalizing, dauser2020relxill} was used to model the complex relativistically blurred reflection component. It offers two variations, the lamppost geometry (\textit{relxilllp}, \textit{relxilllpCp}) and the coronal geometry (\textit{relxill}, \textit{relxillCp}). For our analysis, we use the coronal model for a Comptonization continuum (\textit{relxillCp}, Figure \ref{fig:illust}). It accounts for the gravitationally blurred reflection of the Comptonization continuum from the accretion disk. The spin parameter, $a$, was fixed at 0.998 which is its maximal allowed value. The two indices were tied together while the break ($R_{br}$) and outer ($R_{out}$) radii were fixed at $300R_g$ and $400 R_g$, respectively. The calculations were done for a fixed accretion disk density of $10^{15} cm^{-3}$ and a coronal electron temperature of $60$ keV. The corresponding X-ray fitting plot is shown in Figure \ref{fig:fig_relx}. The power law index for the incident spectrum ($\Gamma_{relx}$) obtained was $\sim2.20$ while the reflection fraction was $\sim9.49$. Ionization parameter ($log\xi$) returned a value of $\sim2.70$. Inner radius ($R_{in}$) lies between $1-1.1 R_{ISCO}$. Emissivity profile is defined by $r^{-Index1}$ and $r^{-Index2}$ between $R_{in}$ \& $R_{br}$ and $R_{br}$ \& $R_{out}$, respectively. These two parameters were tied together. We obtained a lower limit of $8.30$ for Index1, the parameter has an upper limit of $10$ in the model. This agrees with the steep emissivity profiles, with index values greater than $5$, previously been observed for Mrk 335 \citep{gallo2013blurred, wilkins2015flaring, gallo2019evidence}. Table \ref{tab:tab_relx} lists all the parameters of fit.

\begin{table}
    \centering
    \caption{Spectral parameters for Model-I in the energy range $0.3-10.0$ keV.}
    \label{tab:tab_pcf}
    \begin{tabular}{c c c}
        \hline
        \hline
        Model & Parameter & Value \\
        \hline
        zedge       & E (keV) & $0.42_{-0.005}^{+0.006}$ \\
                    & $\tau$ & $0.16_{-0.02}^{+0.02}$ \\
        zpcfabs     & $N_{H,pc}$ ($\times10^{22}$) & $45.77_{-6.06}^{+7.05}$ \\
                    & $f_{c}$  & $0.46_{-0.03}^{+0.03}$ \\
        powerlaw    & $\Gamma_{pl}$ & $2.31_{-0.01}^{+0.01}$ \\
                    & $N_{pl}$($\times10^{-2}$) & $1.80_{-0.10}^{+0.11}$ \\
        zgauss1     & $E_{l}$(keV) & $0.10$ (frozen) \\
                    & N & $0.15_{-0.01}^{+0.01}$ \\
        zgauss2     & $E_{l}$(keV) & $0.76_{-0.01}^{+0.01}$ \\
                    & N($\times10^{-3}$) & $3.23_{-0.26}^{+0.30}$ \\
        zgauss3     & $E_{l}$(keV) & $6.41_{-0.03}^{+0.03}$ \\
                    & $\sigma$(keV) & $<0.12$ \\
                    & N($\times10^{-5}$) & $1.16_{-0.34}^{+0.38}$ \\
        zgauss4     & $E_{l}$(keV) & $7.17_{-0.14}^{+0.17}$ \\
                    & $\sigma$(keV) & $0.25$ (frozen) \\
                    & N($\times10^{-5}$)  & $1.16_{-0.44}^{+0.43}$ \\
        $\chi^2/dof$&   & $302.3/163$\\
        \hline 
    \end{tabular}
\end{table}

\begin{table}
    \centering
    \caption{Spectral parameters for Model-II in the energy range $0.3-10.0$ keV.}
    \label{tab:tab_nthc}
    \begin{tabular}{c c c}
    \hline
    \hline
    Component & Parameter & Value \\
    \hline
    zedge1  & $E_c$ (keV)   & $0.38_{-0.02}^{+0.01}$\\
            & $\tau$        & $0.17_{-0.058}^{+0.098}$\\
    zedge2  & $E_c$ (keV)   & $1.12_{-0.03}^{+0.03}$\\
            & $\tau$        & $0.029_{-0.009}^{+0.008}$\\
    zgauss1 & $E_l$ (keV)   & $6.41_{-0.03}^{+0.03}$\\
            & $\sigma$      & $0.1 (frozen)$\\
            & $N(\times10^{-5})$& $1.68_{-0.217}^{+0.220}$\\
    zgauss2 & $E_l$ (keV)   & $6.96_{-0.04}^{+0.04}$\\
            & $\sigma$      & $0.1 (frozen)$\\
            & $N(\times10^{-5})$& $1.00_{-0.207}^{+0.208}$\\
    simpl   & $\Gamma_{simpl}$      & $1.82_{-0.15}^{+0.03}$\\
            & $f$    & $0.10_{-0.04}^{+0.02}$\\
            & $UpScOnly$    & $1.0 (frozen)$\\
    nthComp & $\Gamma_{nthC}$      & $3.21_{-0.05}^{+0.05}$\\
            & $kT_e$ (keV)  & $>0.69$\\
            & $kT_{bb}$ (keV) & $0.0598_{-0.005}^{+0.003}$\\
            & $N_{nthC}(\times10^{-3})$& $7.62_{-0.57}^{+1.38}$\\
    $\chi^2/dof$&           & $228.52/163$\\
    \hline
    \end{tabular}
\end{table}

\begin{table}
    \centering
    \caption{Spectral parameters for Model-III in the energy range $0.3-10.0$ keV.}
    \label{tab:tab_relx}
    \begin{tabular}{c c c}
    \hline
    \hline
    Component & Parameter & Value \\
    \hline
    zedge1  & $E_c (keV)$   &$0.62_{-0.03}^{+0.02}$\\
            & $\tau$        &$0.18_{-0.04}^{+0.05}$\\
    zedge2  & $E_c (keV)$   &$1.47_{-0.04}^{+0.04}$\\
            & $\tau$        &$0.11_{-0.02}^{+0.02}$\\
    relxillCp & $i (deg)$    &$74.11_{-2.49}^{+1.20}$\\
            & $R_{in} (R_{ISCO})$    &$<1.1$\\
            & $Index1$    &$>8.30$\\
            &$\Gamma_{relx}$     &$2.20_{-0.01}^{+0.01}$\\
            &$log\xi$     &$2.70_{-0.06}^{+0.06}$\\
            &$A_{Fe}$     &$0.67_{-0.07}^{+0.07}$\\
            &$R_{f}$   &$9.495_{-1.39}^{+3.01}$\\
            &$N_{relx}(\times10^{-5})$&$5.597_{-0.62}^{+0.40}$\\
    zgauss1 & $E_l (keV)$   &$6.41_{-0.03}^{+0.03}$\\
            & $\sigma$      &$0.1 (frozen)$\\
            & $N(\times10^{-5})$& $1.43_{-0.252}^{+0.241}$\\
    zgauss2 & $E_l (keV)$   &$6.96_{-0.04}^{+0.05}$\\
            & $\sigma$      &$0.1 (frozen)$\\
            & $N(\times10^{-6})$& $8.18_{-2.45}^{+2.23}$\\
    $\chi^2/dof$&           & $167.6/161$\\
    \hline
    \end{tabular}
\end{table}

\section{Fractional Variability and RMS Spectra} \label{sec:Fvar}

Fractional variability is a useful tool for studying the variability properties of AGNs in different energy ranges. The statistical quantity, fractional root mean square variability amplitude, $F_{var}$, is defined as the square root of normalised excess variance \citep{edelson1990broad, rodriguez1997steps}. Excess variance is an indicator of the `intrinsic' variability of the source in the sense that it is calculated after removing the effect of uncertainties \citep{edelson2002x}. 

Fractional rms variability amplitude was determined following the method explained in \citet{vaughan2003characterizing}. $F_{var}$ is defined as given by equation \ref{eqn:fvar} (Eqn. 10 of \citet{vaughan2003characterizing}) : 
\begin{equation}
    F_{var} = \sqrt{\frac{S^2 - \overline{\sigma_{err}^2}}{\bar{x}^2}}
    \label{eqn:fvar}
\end{equation}
where $S$ is the sample variance given by equation \ref{eqn:samplvar} (Eqn. 6 of \citet{vaughan2003characterizing}) : 
\begin{equation}
    S^2 = \frac{1}{N-1}\sum_{i=1}^{N}(x_i - \bar{x})^2
    \label{eqn:samplvar}
\end{equation}
and $\overline{\sigma_{err}^2}$ is the mean square error as given by equation \ref{eqn:mserr} (Eqn. 9 of \citet{vaughan2003characterizing}) : 
\begin{equation}
    \overline{\sigma_{err}^2} = \frac{1}{N}\sum_{i=1}^{N}\sigma_{err,i}^2
    \label{eqn:mserr}
\end{equation}
The error on $F_{var}$ is given by Eqn. B2 of \citet{vaughan2003characterizing}.

\begin{figure}
    \centering
    \includegraphics[width=0.9\linewidth]{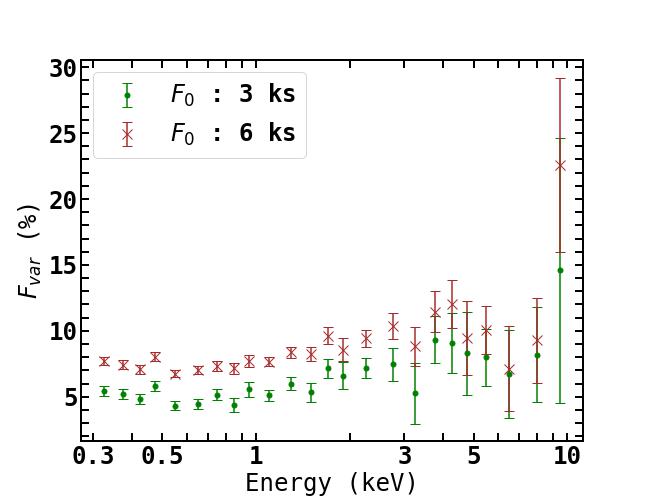}
    \caption{RMS spectra plotting the observed $F_{var}$ as a function of energy. Source lightcurves were extracted at different energy bands. From each of those lightcurves, $F_{var}$ were determined and plotted against energy. This was repeated for lightcurves with different segment sizes, 3 and 6 ks.}
    \label{fig:fig_rmsspec}
\end{figure}

We extracted light curves in multiple energy bins. For each lightcurve we consider segments of different lengths, giving us segments of sizes, 500 secs, 700 secs, 1 ks, 1.5 ks, 3 ks and 6 ks. For each case average count, variance and excess variance were calculated at all energy bins, from which $F_{var}$ and its errors were determined using the above equations. We denote the $F_{var}$ values, thus estimated from the observation, by $F_0$. The rms spectra for this observation was then created by plotting $F_0$ as a function of energy. We find that for segment sizes lower than 3 ks, for some of the energy bins, the excess variances were not detected. Hence we limit the analysis to segments of sizes $3.0$ and $6.0$ ks. Figure \ref{fig:fig_rmsspec} plots the rms spectra for the 3 and 6 ks segments.

\section{Fractional Variability Prediction} \label{sec:Fpred}

\subsection{Method} \label{sec:method}
We ascribe the variation of the count rate at a certain energy band, $R_i$ (where the subscript $i$ denotes the energy band) being caused by variations in the values of two spectral parameters, $X$ and $Y$. In the linear approximation this implies that the variation of the count rate, $\Delta R_i$ can be written as :
\begin{equation}
  \ \Delta R_i(X,Y) = \pdv{R_i}{X}_{X_0}\Delta X + \pdv{R_i}{Y}_{Y_0}\Delta Y
    \label{eqn:Ri_eqn}
\end{equation}
where $X_0$ and $Y_0$ are the steady state values of $X$ and $Y$ and $\Delta X$ and $\Delta Y$ are their variation. Noting the predicted normalised $F_{var}$ at the energy band denoted by $i$ is $F_{Pi}$, given by, $F_{Pi}^2 = (\Delta R/R_i)^2$, this implies that :
\begin{equation}
    F_{Pi}^2 = A_i^2 \delta X^2 + B_i^2 \delta Y^2 + 2 A_iB _i C_{XY} |\delta X| |\delta Y|
    \label{eqn:fp}
\end{equation}
where $\delta X = \Delta X/X_0$, $\delta Y = \Delta Y/Y_0$, $A_i =\frac{X_0}{R_i} \pdv{R_i}{X}_{X_0}$ and $B_i =\frac{Y_0}{R_i} \pdv{R_i}{Y}_{Y_0}$. $C_{XY}$ represents both the strength and the sign of the cross term, such that $C_{XY} = 1,0,-1$ represent the cases when the $\delta X$ and $\delta Y$ are correlated, un-correlated and anti-correlated. Fractional values of $C_{XY}$ can be either interpreted as when $\delta X$ and $\delta Y$ are partially correlated or that there is a phase lag between them such that $C_{XY} = cos\phi$.

The partial derivatives $ \pdv{R_i}{X}_{X_0}$ and $ \pdv{R_i}{Y}_{Y_0}$ were estimated numerically using the \textsc{xspec} software. We considered the best fit model spectra and then found the change in the model count rate at a given energy band by introducing a $10\%$ variation in the value of the parameter. The ratio of this change in the count rate to that of the parameter value was taken as the partial derivative. Note that since this exercise was undertaken in the model count rates (and not the photon rate), the response of the instrument is being taken into account. The best fit spectra values were then used to estimate $A_i$ and $B_i$. 

The predicted $F_p$ was then compared with the observed ones to obtain the best fit values for $|\delta X|$, $|\delta Y|$ and $C_{XY}$ by minimising the chi-square function :
\begin{equation}
    \chi^2 = \sum_{i=0}^n \left(\frac{F_{0i}^2 - F_{Pi}^2}{\sigma_i}\right)^2
    \label{eqn:chisq}
\end{equation}
Here, $F_{0i}^2$ is the squared observed fractional rms variability amplitude ($F_{var}$) and $\sigma_i$ is the error on $F_{0i}^2$. We note that the minimisation of $\chi^2$ can be done analytically and one can obtain expressions for the best fit values as shown in \ref{sec:appa}. However, these analytical expressions sometimes gave unphysical best fit values such as $|\delta X|$ or $|\delta Y| < 0$ or $|C_{XY}| > 1$. Thus, we had to use the python library \textsc{lmfit} \citep{newville2016lmfit}, to minimise the $\chi^2$ which allows for the parameter values to be within a specified range.

\subsection{Model-I : Partial Covering Model}
To model the intrinsic variability, we consider the different spectral parameters. For a given segment size, the parameters were varied in combinations of one and two and all different combinations were considered. For each combination, the parameter values were varied slightly and the corresponding $R_{0i}$, $R_{1i}$ and $R_{2i}$ values were determined for all different energy bins. From these, $A_i$ and $B_i$ can be estimated as explained above. These values were then input into \textsc{lmfit} along with corresponding $F_{0i}^2$ and $\sigma_i$ values. The fit parameters $\delta X^2$, $\delta Y^2$, $C_{XY}$ and the result $F_P$ were obtained as output. The combination that gives the best $\chi^2$ fit was determined, under the conditions that neither $\delta X^2$ nor $\delta Y^2$ can be negative and $C_{XY}$ lies between -1 and +1. The predicted rms spectra for a given segment size and a given combination of model parameters can be created by plotting $F_P$ as a function of energy.

The spectral parameters were varied individually and in pairs. For both segment sizes, the best fit was given by a combination of the covering fraction ($f_c$) and the power law photon index ($\Gamma_{pl}$). For a segment length of 3 ks, the combination is obtained for $\delta f_c$ of $5.93\%$ and $\delta \Gamma_{pl}$ of $\sim2.14\%$. Similarly, for the 6 ks case, the lowest $\chi^2$ is obtained again for $\delta f_c$ and $\delta \Gamma_{pl}$ values of $8.86\%$ and $2.43\%$, respectively. In both the cases, the two parameters are slightly correlated with $C_{XY}$ values of $\sim 0.3$. Table \ref{tab:par_comb_pcf} lists the parameter combinations with lowest $\chi^2$ for three cases, all other combinations give higher values of $\chi^2$. For both the 3ks and 6ks cases, we plot a continuous curve by determining $F_P$ values at very small energy bins (0.1 keV) using these fit parameter values. The resultant spectra are plotted in Figures \ref{fig:fp_3_pcf} and \ref{fig:fp_6_pcf}.

\subsection{Model-II : Thermal Comptonization}
For the thermal Comptonization model, we consider all possible parameters that determine the X-ray spectrum in the full energy range. As with the partial covering model, the parameters are varied individually and in combinations of two. We found that varying the electron temperature ($kT_e$) does not bring about a significant change in the count rates and hence was not considered during the analysis. The whole procedure was repeated for all combinations of the model parameters for both segment sizes. For the 3 ks case, we get the best fit by varying the parameters, photon index ($\Gamma_{nthC}$) and normalisation ($N_{nthC}$) of the warm corona. The values of $\delta \Gamma_{nthC}$ and $\delta N_{nthC}$ are $\sim2.49\%$ and $\sim5.66\%$ respectively with a $C_{XY}$ $\sim-0.59$, indicating a significant anti-correlation between the two parameters. For the case with a larger segment of 6 ks length, the best fit is again given by the combination of photon index ($\Gamma_{nthC}$) and normalisation ($N_{nthC}$), with the values of $\delta \Gamma_{nthC}$ and $\delta N_{nthC}$ being $\sim2.87\%$ and $\sim8.15\%$. In this case also, we see a slight correlation between the two parameters with $C_{XY}$ $\sim-0.51$. Parameter combinations with the least values of $\chi^2$ are listed in Table \ref{tab:par_comb_nthc}. The continuous curve is plotted for small energy bins of size $0.1$ keV, as shown in Figures \ref{fig:fp_3_nthc} and \ref{fig:fp_6_nthc}.

\subsection{Model-III : Relativistic Reflection}
\begin{figure*}
    \centering
    \subfloat{%
        \includegraphics[width=0.48\linewidth]{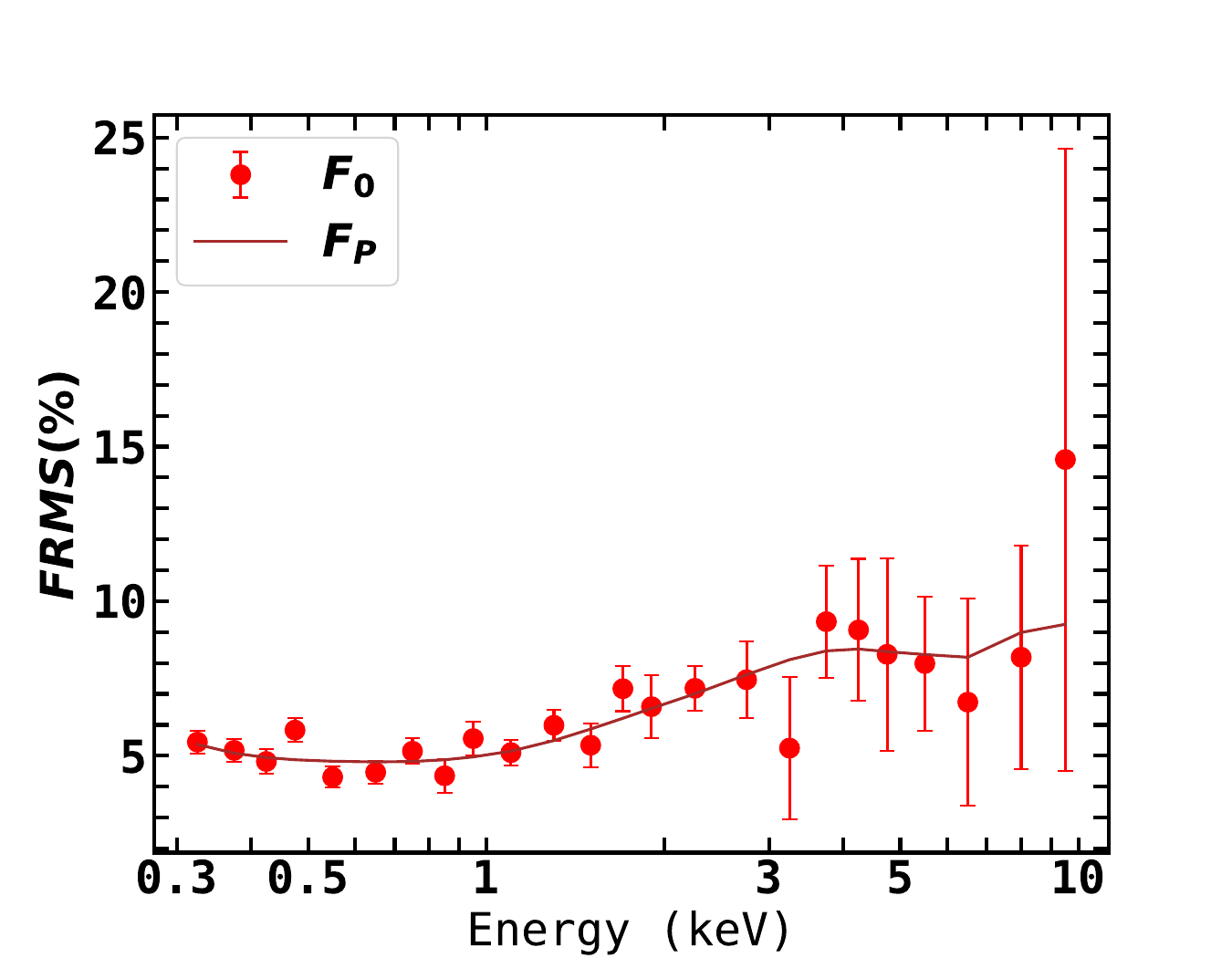}%
        \label{fig:fp_3_pcf}%
    }\qquad
    \subfloat{%
        \includegraphics[width=0.48\linewidth]{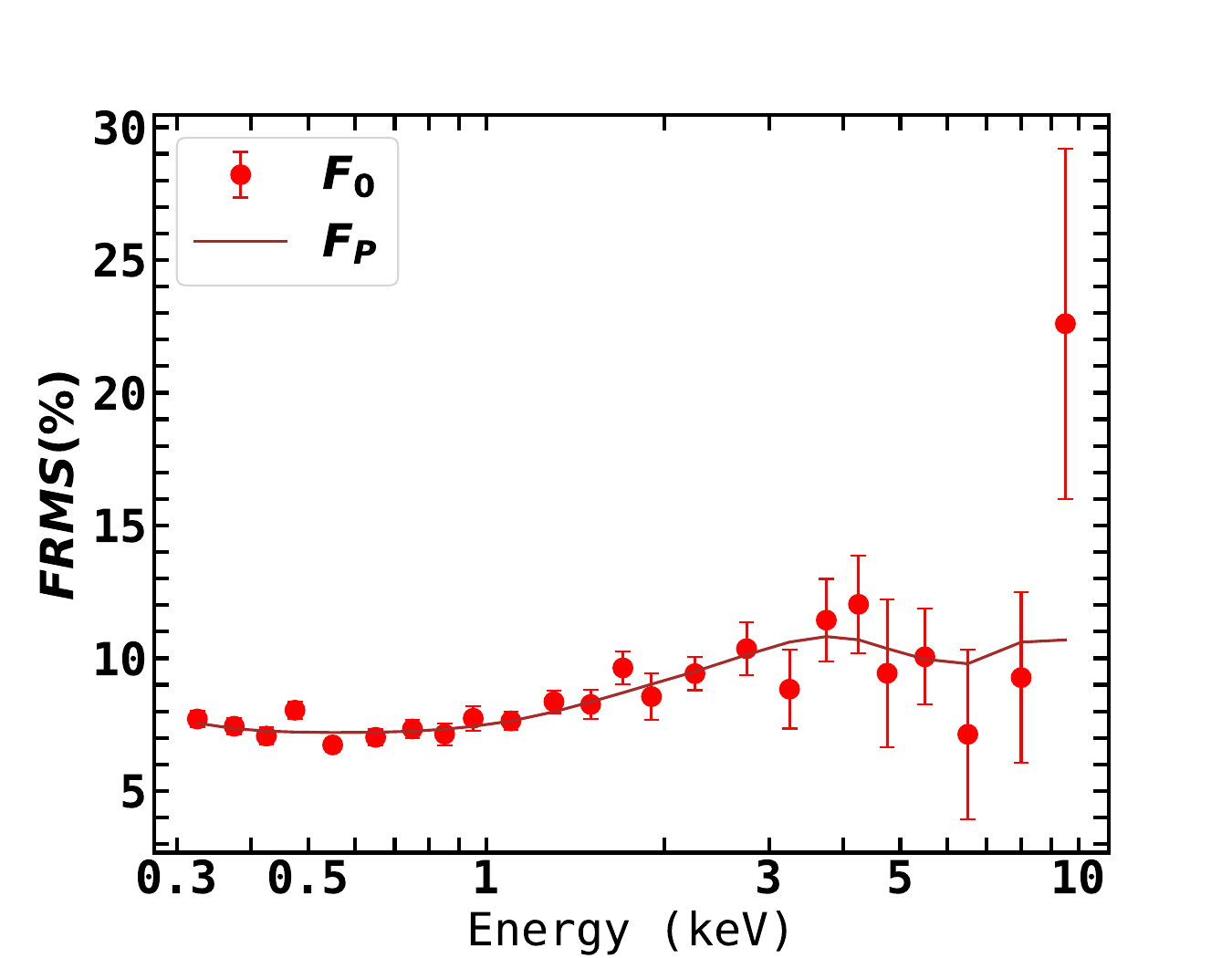}%
        \label{fig:fp_6_pcf}%
    }\qquad
    \subfloat{%
        \includegraphics[width=0.48\linewidth]{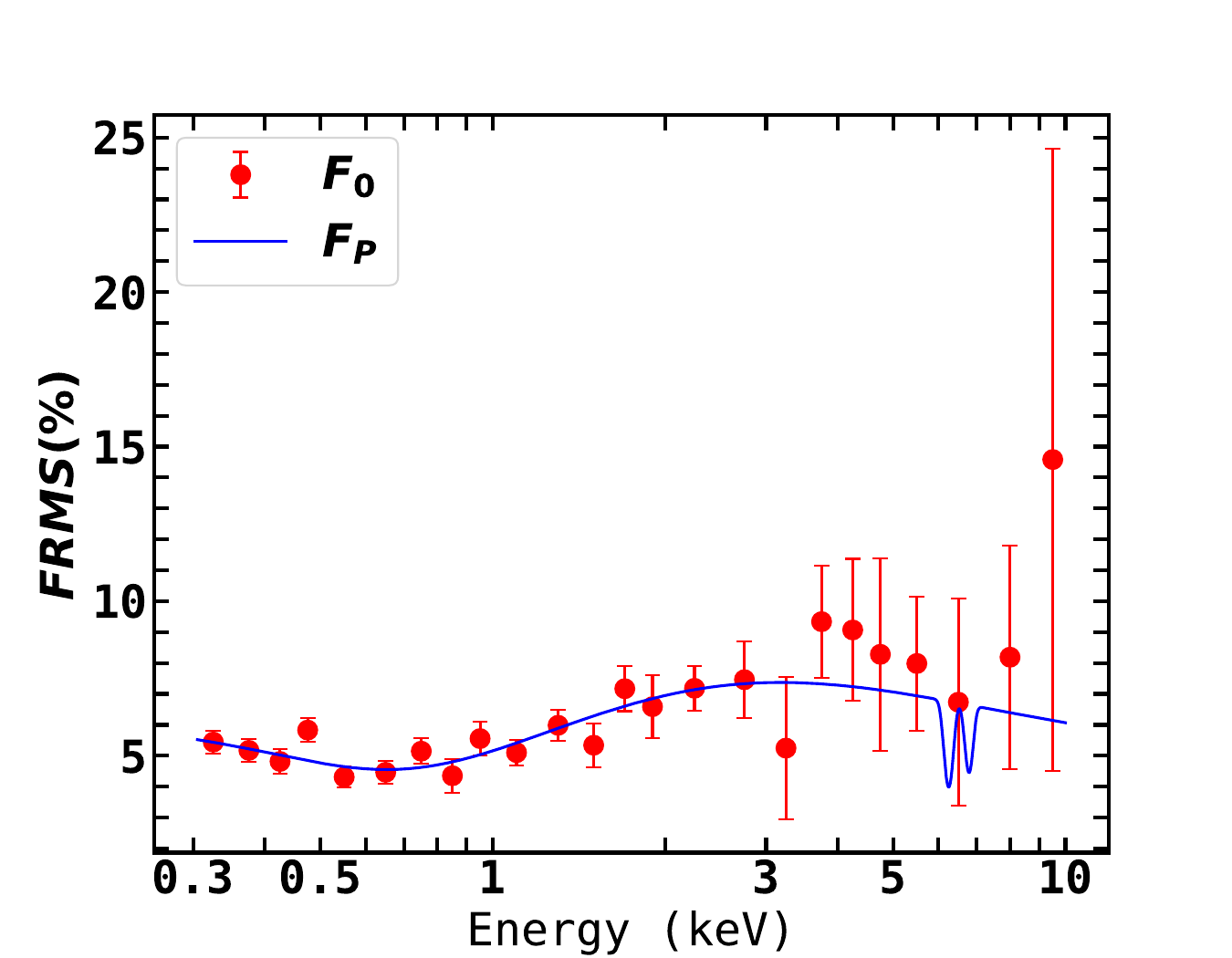}%
        \label{fig:fp_3_nthc}%
    }\qquad
    \subfloat{%
        \includegraphics[width=0.48\linewidth]{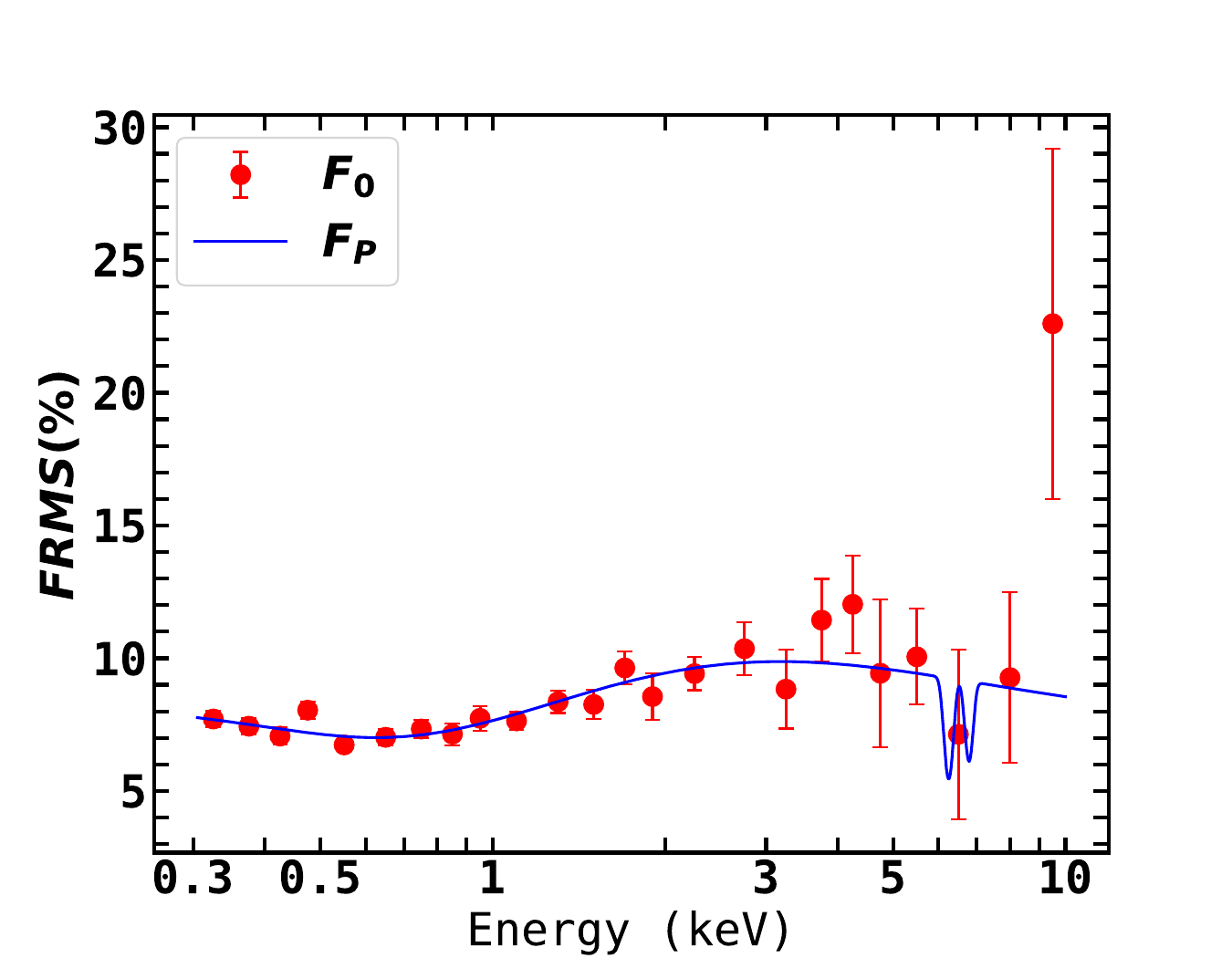}%
        \label{fig:fp_6_nthc}%
    }\qquad
    \subfloat{%
        \includegraphics[width=0.48\linewidth]{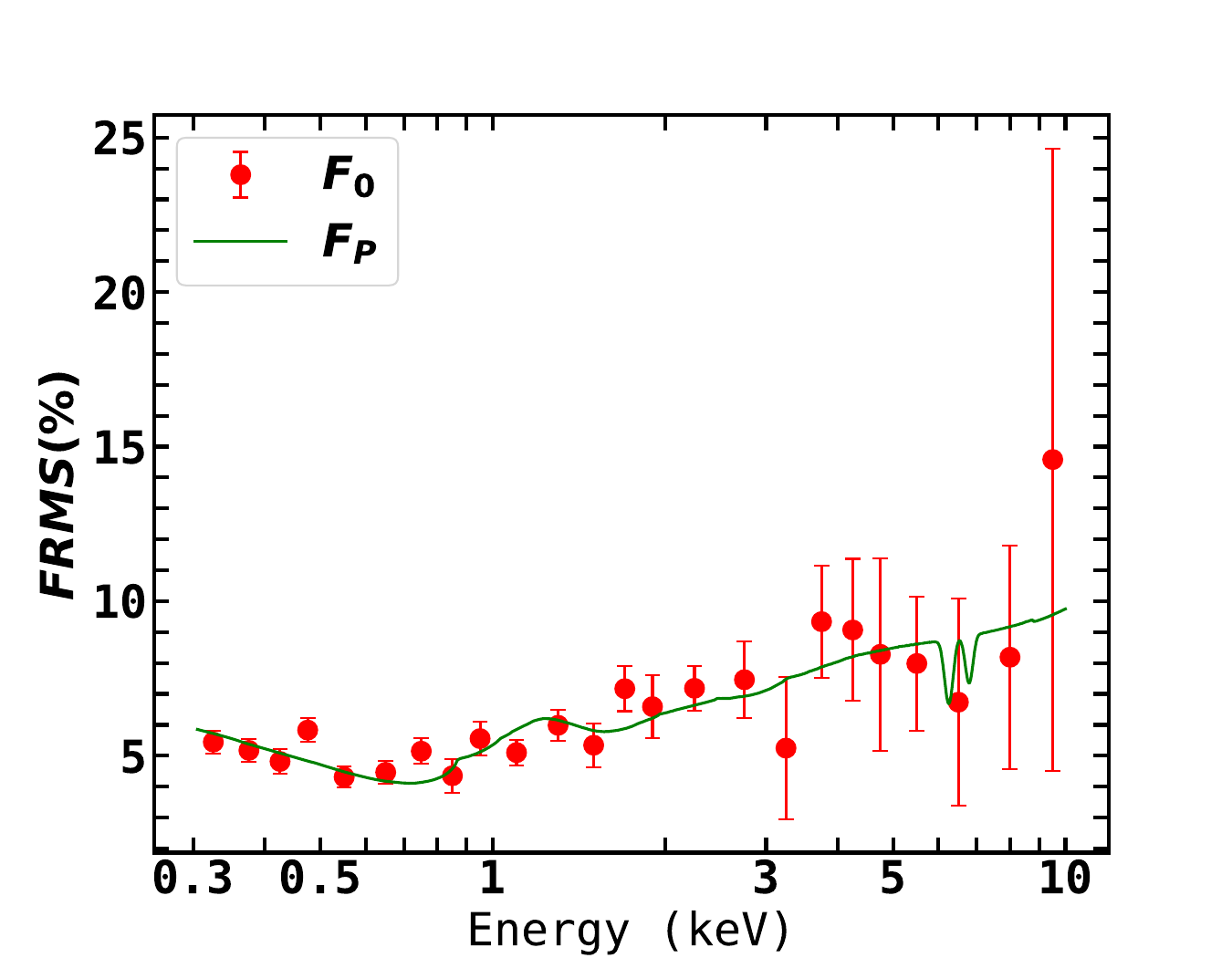}%
        \label{fig:fp_3_relx}%
    }\qquad
    \subfloat{%
        \includegraphics[width=0.48\linewidth]{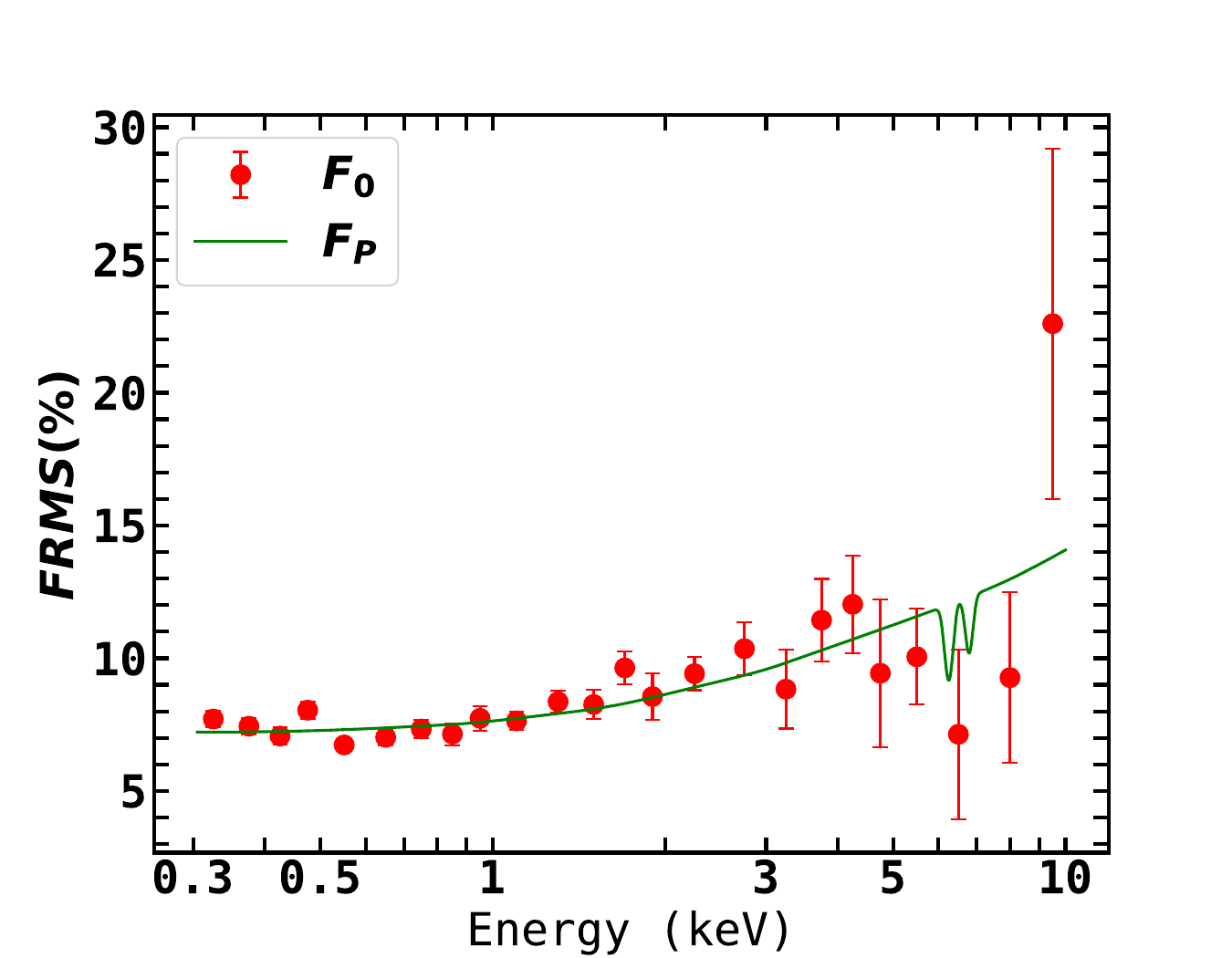}%
        \label{fig:fp_6_relx}%
    }
    \caption{The observed $F_{var}$ values ($F_0$) are plotted along with a continuous curve corresponding to the predicted $F_{var}$ ($F_P$) using all the three models. Left and right panels show the plots for segment sizes 3 ks and 6 ks, respectively. From the top, $F_P$ is predicted using the partial covering model (brown curve), warm Comptonization (blue curve) and blurred reflection (green curve), respectively.}
    \label{fig:fig4}
\end{figure*}

As was done with the previous models, here we consider all parameters that determine the spectrum, except the iron abundance. We find that the model count rates are not sensitive to changes in the inner disk radius ($R_{in}$) and hence was not considered for the analysis. The procedure as described earlier was followed and the rms spectrum was predicted. The power law photon index ($\Gamma_{relx}$) and the accretion disk's ionisation parameter ($log\xi$) of \textit{relxillCp} gives the best fit for the case of $3$ ks long segment. Here the values of $\delta \Gamma_{relx}$ and $\delta log\xi$ obtained are $\sim1.91\%$ and $\sim2.64\%$ and the two have a $C_{XY}$ value of $\sim-0.49$. For the $6$ ks case, the best fit is given by a different parameter combination, power law index ($\Gamma_{relx}$) and normalisation ($N_{relx}$) parameters of \textit{relxillCp}. We saw that, in this case, the two parameters are anti-correlated with the values of $\delta \Gamma_{relx}$ and $\delta N_{relx}$ being $\sim1.53\%$ and $\sim8.90\%$ and $C_{XY}<-0.53$. Parameter combinations giving the best $\chi^2$ values are tabulated in Table \ref{tab:par_comb_relx}. As before, a continuous curve was also plotted by choosing very small energy bins of $0.1$ keV size. The rms spectra are shown in Figures \ref{fig:fp_3_relx} and \ref{fig:fp_6_relx}.

\begin{table}
    \centering
    \caption{Fit statistics for different parameter combinations of Model-I.}
    \label{tab:par_comb_pcf}
    \resizebox{\linewidth}{!}{
    \begin{tabular}{c c c c c c c}
    \hline\hline
    Segment & $X$ & $\delta X$ & $Y$ & $\delta Y$ & $C_{XY}$ & $\chi^2/dof$\\
    Size & & $(\%)$ & & $(\%)$ & &\\
    \hline
            & $f_{c}$ & $5.93_{-0.22}^{+0.20}$ & $\Gamma_{pl}$ & $2.14_{-0.25}^{+0.23}$ & $0.29_{-0.06}^{+0.06}$ & $17.89/21$ \\
        3 ks & $f_{c}$ & $13.14$ & $N_{pl}$ & $14.39$ & $0.91$ & $22.15/21$ \\
            & $\Gamma_{pl}$ & $1.97$ & $N_{pl}$ & $6.30$ & $0.35$ & $25.47/21$ \\
    \hline
            & $f_{c}$ & $8.86_{-0.16}^{+0.18}$ & $\Gamma_{pl}$ & $2.43_{-0.26}^{+0.22}$ & $0.28_{-0.04}^{+0.05}$ & $18.95/21$ \\
        6 ks & $f_{c}$ & $15.47$ & $N_{pl}$ & $16.24$ & $0.84$ & $24.82/21$ \\
            & $\Gamma_{pl}$ & $2.50$ & $N_{pl}$ & $9.37$ & $0.45$ & $39.83/21$ \\
    \hline
    \end{tabular}
    }
\end{table}

\begin{table}
    \caption{Fit statistics for different parameter combinations of Model-II.}
    \label{tab:par_comb_nthc}
    \resizebox{\linewidth}{!}{
    \begin{tabular}{c c c c c c c}
    \hline\hline
    Segment & $X$ & $\delta X$ & $Y$ & $\delta Y$ & $C_{XY}$ & $\chi^2/dof$\\
    Size & & $(\%)$ & & $(\%)$ & &\\
    \hline
        & $\Gamma_{nthC}$ & $2.49_{-0.29}^{+0.26}$ & $N_{nthC}$ & $5.66_{-0.19}^{+0.19}$ & $-0.59_{-0.04}^{+0.04}$ & $16.93/21$ \\
        3 ks & $f$ & $10.40$ & $N_{nthC}$ & $4.91$ & $0.04$ & $21.88/21$ \\
        & $\Gamma_{simpl}$ & $2.29$ & $N_{nthC}$ & $5.37$ & $-1.00$ & $36.85/21$ \\
    \hline
        & $\Gamma_{nthC}$ & $2.87_{-0.30}^{+0.23}$ & $N_{nthC}$ & $8.15_{-0.16}^{+0.16}$ & $-0.51_{-0.03}^{+0.03}$ & $16.98/21$ \\
        6 ks & $f$ & $10.21$ & $N_{nthC}$ & $7.34$ & $0.17$ & $24.70/21$ \\
        & $\Gamma_{simpl}$ & $2.25$ & $N_{nthC}$ & $7.72$ & $-0.81$ & $53.34/21$ \\
    \hline
    \end{tabular}
    }
\end{table}
\begin{table}
    \centering
    \caption{Fit statistics for different parameter combinations of Model-III.}
    \label{tab:par_comb_relx}
    \resizebox{\linewidth}{!}{
    \begin{tabular}{c c c c c c c}
    \hline\hline
    Segment & $X$ & $\delta X$ & $Y$ & $\delta Y$ & $C_{XY}$ & $\chi^2/dof$\\
    Size & & $(\%)$ & & $(\%)$ & &\\
    \hline
        & $\Gamma_{relx}$ & $1.91_{-0.17}^{+0.12}$ & $log\xi$ & $2.64_{-0.14}^{+0.10}$ & $-0.49_{-0.08}^{+0.10}$ & $22.81/21$\\
     3 ks & $\Gamma$ & $1.58$ & $N_{relx}$ & $6.50$ & $-0.65$ & $23.91/21$\\
        & $R_{f}$ & $6.57$ & $N_{relx}$ & $8.89$ & $-0.92$ & $30.93/21$\\
    \hline
        & $\Gamma_{relx}$ & $1.53_{-0.33}^{+0.23}$ & $N_{relx}$ & $8.90_{-0.24}^{+0.21}$ & $<-0.53$ & $29.86/21$\\
     6 ks & $log\xi$ & $15.82$ & $N_{relx}$ & $24.44$ & $-0.95$ & $44.57/21$\\
        & $R_{f}$ & $7.44$ & $N_{relx}$ & $12.74$ & $-1.00$ & $46.76/21$\\
    \hline
    \end{tabular}
    }
\end{table}

\section{Results and Discussion} \label{sec:result}

Mrk 335 has an X-ray spectrum resembling that of a typical Seyfert-1 galaxy with a strong soft excess, with significant flux variations in the X-ray band. Broadband X-ray spectrum of the source can be interpreted as a partial covering absorber modifying the power law, a warm thermal Comptonization component or as the gravitational blurred reflection off the accretion disk. We fit spectral models representing these interpretations to the \textit{XMM-Newton} observation of Mrk 335 and find that the spectrum can be well represented by all the aforementioned models. The photons indices obtained for all three cases are well within the range of 1.8 - 2.5, typically observed for such galaxies \citep{nandra1994ginga, page2005xmm}.

The \textit{XMM-Newton} observation considered in this work has been reported to show high variability in its X-ray emission \citep{wilkins2015driving, mastroserio2020multi}. We estimated the fractional rms amplitude, $F_{var}$, as a function of energy, $E$, for two different segments of lengths $3$ ks and $6$ ks. We developed a scheme which, for a given spectral model fit, would predict $F_{var} (E)$ as being caused by the coherent variation of two of the spectral parameters, enabling the identification of the spectral parameters responsible for the observed variability. As can be seen from Figures \ref{fig:fig_rmsspec} and \ref{fig:fig4}, $F_{var}$ is not a constant with energy. There is evidence that $F_{var}$ decreases from $0.3$ keV to about $0.7$ keV and rises for higher energies. Our analysis shows that none of the three models can explain the observed profile using a single parameter variation; which always results in a reduced $\chi^2$ greater than $45/21$.

The partial covering interpretation of the X-ray spectrum, involves a single power law emission along with a neutral patchy absorber that partially covers the source. To explain the observed variation of fractional rms with energy, the analysis undertaken here shows that a variation in the primary emission itself is warranted. This is manifested as the best fit requires a variation in the power law index ($\Gamma_{pl}$). A previous study by \citet{yamasaki2016origin} on IRAS 3224- 3809 shows that, rms variability over a time scale of $\sim$ 1 day is explained by changing the partial covering fraction. From our analysis of Mrk 335 using the partial covering scenario, variations on smaller time scale is again explained by variations in the covering fraction ($f_c$) along with the photon index. The results are the same for both the segment sizes, of 3 ks and 6 ks, with similar values of $\chi^2$. In both the cases, the parameters are slightly correlated with a $C_{XY}$ of $\sim 0.3$.

In the warm Comptonization interpretation, a hot corona Comptonizes photons from the warm corona, with the latter giving rise to the soft excess. Thus a variation of the normalization of the warm Comptonization component (with other parameters being constant) will lead to an overall normalization change in the full spectrum, leading to $F_{var}$ which is constant with energy. In this case, the rms analysis shows that there needs to be a variation of photon index of the warm Comptonization ($\Gamma_{nthC}$). Moreover, the variation of the normalization and $\Gamma_{nthC}$ are anti-correlated with each other since the cross-correlation $C_{XY}$ turns out to be negative. Note that $\Gamma_{nthC}$ depends on the optical depth and temperature of the warm corona. A decrease in $\Gamma_{nthC}$ implies an increase in the warm corona temperature or optical depth. Thus, the analysis suggests that the energy dependent $F_{var}$ for both time-scales can be explained in terms of variations of the normalization and temperature (or optical depth) of the warm corona. It is interesting to note that although there is significant variability at high energies, the energy dependent $F_{var}$ can be explained solely due to variations of the warm corona parameters, which changes the seed photons of the hot Comptonization component.

For the case when the soft excess is explained by the reflection component, we note the exceptionally high values of the emissivity index and the reflection fraction parameter. For this spectrum, a variation in the normalization of the reflection component will give rise to a constant energy independent $F_{var}$. The r.m.s. analysis shows that if combined with a change in the photon index $\Gamma_{relx}$, this leads to an increase in $F_{var}$ with energy which can broadly explain the data as shown in Figure \ref{fig:fp_6_relx} (segment size of 6 ks). However, this does not produce the dip in $F_{var}$ at $\sim0.7$ keV, leading to a higher value of $\chi^2/dof = 30/21$ (Table \ref{tab:par_comb_relx}). It is interesting to note that a combination of variations in photon index $\Gamma_{relx}$ and ionization parameter $\log \xi$ can produce such a dip as seen for the fitting for segment size of 3 ks (Figure \ref{fig:fp_3_relx}). The same combination does not provide a good fit when the segment size is 6 ks resulting in $\chi^2/dof = 47.92/21$. This suggests that perhaps more than two parameter variation are required to explain the data. The steep emissivity profile and an extreme value of the reflection parameter for a non-point-like X-ray source, combined with a requirement for variations in more than two parameters, indicates that the blurred reflection model might not be a suitable description of the Mrk 335 spectrum.

We note that the fit statistics are somewhat low for a few of these combinations, especially in the partial covering and the Comptonization models. Such low values of $\chi^2$ suggests an over estimation of errors. This could possibly arise when there is an intrinsic correlation between the variabilities in different energy bands \citep{ingram2019error}. They stipulate the use of new formulae for calculating the errors in such situations, assuming that the variability power spectrum is known. However in our current case, the data is not sufficiently long enough for such detailed analysis. Furthermore it’s not clear how to take this effect into account when the $F_{var}$ is being computed directly from the lightcurve rather than power spectrum analysis.

The method prescribed here, attributes the X-ray variability to be caused by two spectral parameters varying in combination. It may also be possible that the system is indeed more complex and more than two parameters maybe varying. More detailed modelling taking into account the physical dependence of the spectral parameters on each other is required to make more concrete conclusions. Moreover, the results depend on the specific spectral model used to fit the energy spectrum. More sophisticated spectral models such as those which compute the emissivity index self consistently for a specific geometry (e.g. the \textit{relxilllp}, for the `lamp-post' model) and incorporating complex absorption models to account for the observed edges may be required. Nevertheless, the method presented in this work, has the potential to differentiate between spectrally degenerate models. Moreover, the work needs to be extended to other data sets especially at high energies (perhaps using \textit{NuStar} observations) to complement the results presented here.

\section*{Acknowledgements}
Authors KA and KJ acknowledge the financial support from ISRO (Sanction Order:No.DS\_2B-13013(2)/11/2020-Sec.2). KA thank the IUCAA visiting program. KJ and RS acknowledge the associateship program of IUCAA, Pune. This research has made use of data, software and/or web tools obtained from the High Energy Astrophysics Science Archive Research Center (HEASARC), a service of the Astrophysics Science Division at NASA/GSFC and of the Smithsonian Astrophysical Observatory's High Energy Astrophysics Division.

\appendix

\section{Analytical solution for fit parameters of predicted fractional variability}
\label{sec:appa}
The parameters $\delta X^2$, $\delta Y^2$ and $C_{XY}$ in Eqn. \ref{eqn:fp} may also be determined analytically. This makes use of the function $\chi^2$ defined by Eqn. \ref{eqn:chisq}. For the cases,
\begin{subequations}
    \begin{gather}
        \frac{d\chi^2}{d\delta X^2} = 0   \quad  \frac{d\chi^2}{d\delta Y^2} = 0 \quad \frac{d\chi^2}{dC_{XY}} = 0
        \tag{\theequation a-c}
    \end{gather}
\end{subequations}
The set of equations given below follows directly
\begin{subequations}
    \begin{gather}
        \resizebox{0.6\linewidth}{!}{$
        \begin{bmatrix}
            \sum\limits_{i} \frac{A_i^4}{\sigma^2} & \sum\limits_i \frac{A_i^2B_i^2}{\sigma^2} & \sum\limits_i \frac{A_i^3B_i^2}{\sigma^2} \\
            \sum\limits_i \frac{A_i^2B_i^2}{\sigma^2} & \sum\limits_i \frac{B_i^4}{\sigma^2} & \sum\limits_i \frac{A_iB_i^3}{\sigma^2} \\
            \sum\limits_i \frac{A_i^3B_i}{\sigma^2} & \sum\limits_i \frac{A_iB_i^3}{\sigma^2} & \sum\limits_i \frac{A_i^2B_i^2}{\sigma^2} \\
        \end{bmatrix}
        \begin{bmatrix}
            \delta X^2 \\ \delta Y^2 \\ C_{XY}
        \end{bmatrix}
        =
        \begin{bmatrix}
            \sum\limits_i \frac{F_{0i}^2A_i^2}{\sigma^2} \\ \sum\limits_i \frac{F_{0i}^2B_i^2}{\sigma^2} \\ \sum\limits_i \frac{F_{0i}^2A_iB_i}{\sigma^2} \\
        \end{bmatrix}
        $}
        \tag{\theequation a-c}
        \label{eqn:matrixeqn}
    \end{gather}
\end{subequations}

The equations \ref{eqn:matrixeqn} can then be solved for $\delta X$, $\delta Y$ and $C_{XY}$ respectively. If the term $\sum\limits_{i=0}^n \frac{1}{\sigma^2}A_i^2B_i^2$ was denoted as $A_i^2B_i^2$ and so on, then the equations become :

\begin{equation}\label{eqn:delxsq}
    \resizebox{\linewidth}{!}{$
    \delta X^2 = \frac{[(F_{0i}^2A_i^2)(B_i^4)(A_i^2B_i^2) + (F_{0i}^2A_iB_i)(A_i^2B_i^2)(A_iB_i^3) + (F_{0i}^2B_i^2)(A_i^3B_i)(A_iB_i^3)] - [(F_{0i}^2A_i^2)(A_iB_i^3)(A_iB_i^3) + (F_{0i}^2B_i^2)(A_i^2B_i^2)(A_i^2B_i^2) + (F_{0i}^2A_iB_i)(B_i^4)(A_i^3B_i)]}
    {[(A_i^4)(B_i^4)(A_i^2B_i^2) + (A_i^2B_i^2)(A_i^3B_i)(A_iB_i^3) + (A_i^3B_i)(A_i^2B_i^2)(A_iB_i^3)] - [(A_i^4)(A_iB_i^3)(A_iB_i^3) + (A_i^2B_i^2)(A_i^2B_i^2)(A_i^2B_i^2) + (B_i^4)(A_i^3B_i)(A_i^3B_i)]}    
    $}
\end{equation}
\begin{equation}\label{eqn:delysq}
    \resizebox{\linewidth}{!}{$
    \delta Y^2 = \frac{[(F_{0i}^2B_i^2)(A_i^4)(A_i^2B_i^2) + (F_{0i}^2A_i^2)(A_iB_i^3)(A_i^3B_i) + (F_{0i}^2A_iB_i)(A_i^3B_i)(A_i^2B_i^2)] - [(F_{0i}^2A_iB_i)(A_i^4)(A_iB_i^3) + (F_{0i}^2A_i^2)(A_i^2B_i^2)(A_i^2B_i^2) + (F_{0i}^2B_i^2)(A_i^3B_i)(A_i^3B_i)]}
    {[(A_i^4)(B_i^4)(A_i^2B_i^2) + (A_i^2B_i^2)(A_i^3B_i)(A_iB_i^3) + (A_i^3B_i)(A_i^2B_i^2)(A_iB_i^3)] - [(A_i^4)(A_iB_i^3)(A_iB_i^3) + (A_i^2B_i^2)(A_i^2B_i^2)(A_i^2B_i^2) + (B_i^4)(A_i^3B_i)(A_i^3B_i)]}    
    $}
\end{equation}
\begin{equation}\label{eqn:cxy}
    \resizebox{\linewidth}{!}{$
    C_{XY} = \frac{[(F_{0i}^2A_iB_i)(A_i^4B_i^4) + (F_{0i}^2B_i^2)(A_i^2B_i^2)(A_i^3B_i) + (F_{0i}^2A_i^2)(A_i^2B_i^2)(A_i^3B_i)] - [(F_{0i}^2B_i^2)(A_i^4)(A_iB_i^3) + (F_{0i}^2A_iB_i)(A_i^2B_i^2)(A_i^2B_i^2) + (F_{0i}^2A_i^2)(B_i^4)(A_i^3B_i)]}{[(A_i^4)(B_i^4)(A_i^2B_i^2) + (A_i^2B_i^2)(A_i^3B_i)(A_iB_i^3) + (A_i^3B_i)(A_i^2B_i^2)(A_iB_i^3)] - [(A_i^4)(A_iB_i^3)(A_iB_i^3) + (A_i^2B_i^2)(A_i^2B_i^2)(A_i^2B_i^2) + (B_i^4)(A_i^3B_i)(A_i^3B_i)]}
    $}
\end{equation}

For the special case where only one parameter was varied ($\Delta Y=0$), the equations reduce to
\begin{subequations}
    \begin{gather}
        \delta X^2 = \frac{F_{0i}^2A_i^2}{A_i^4} \quad \delta Y^2 = 0 \quad C_{XY}=0
        \tag{\theequation a-c}
    \end{gather}
\end{subequations}
Substituting these values of $\delta X^2$, $\delta Y^2$ and $C_{XY}$ in equation \ref{eqn:fp} will give the value of the predicted $F_P$.

\bibliographystyle{elsarticle-harv} 
\bibliography{main_Revised2}






\end{document}